# Affordances Provide a Fundamental Categorization Principle for Visual Scenes

Michelle R. Greene (1), Christopher Baldassano (1), Andre Esteva (2), Diane M. Beck (3) & Li Fei-Fei (1)

(1) Stanford University, Department of Computer Science
(2) Stanford University, Department of Electrical Engineering
(3) University of Illinois at Urbana-Champaign

Author for correspondence:

Michelle R. Greene
Department of Computer Science
Room 240
Stanford University
353 Serra Mall
Stanford, CA 94305

**Abstract**

How do we know that a kitchen is a kitchen by looking? Relatively little is known about how we conceptualize and categorize different visual environments. Traditional models of visual perception posit that scene categorization is achieved through the recognition of a scene's objects, yet these models cannot account for mounting evidence that human observers are relatively insensitive to the local details in an image. Psychologists have long theorized that the affordances, or the actionable possibilities of a stimulus are pivotal to its perception. To what extent are scene categories created from similar affordances? Using a large-scale experiment using hundreds of scene categories, we show that the activities afforded by a visual scene provide a fundamental categorization principle. Affordance-based similarity explained the majority of the structure in human scene categorization patterns, outperforming alternative similarities based on objects or visual features. When all these models are combined, affordances provide the majority of the predictive power in the combined model, and nearly half of the total explained variance is captured only by affordances. These results challenge many existing models of high-level visual perception, and provide immediately testable hypotheses for the functional organization of the human perceptual system.

**Significance Statement**

How do we know that a kitchen is a kitchen by looking? Models of visual perception assume that scene identification is facilitated through object recognition. However, these models fail to account for observers' relative insensitivity to local image details. We explore an alternative view that posits that a scene's identity is determined by the possibilities for actions that a scene affords (its affordances). In a large-scale experiment using hundreds of scene categories, we found that human scene similarity ratings were more closely related to affordance-based similarity than to object or visual feature-based models. Combining models revealed that nearly half of the explained variance was captured only by affordances. This work demonstrates that affordances provide a fundamental grouping principle for scenes.

**Introduction**

"The question 'What makes things seem alike or different?' is one so fundamental to psychology that very few psychologists have been naïve enough to ask it" (1).

Although more than half a century has passed since Attneave issued this challenge, we still have little understanding of how we categorize and conceptualize visual content. Traditionally, it has been assumed that scenes are categorized according to their component features and objects (2–7). Mounting behavioral evidence, however, indicates that human observers have high sensitivity to the global meaning of an image (8–11), and very little sensitivity to the local objects and features that are outside the focus of attention (12). Consider the image of the kitchen in Figure 1. If scene categories are determined by objects, then we would expect the kitchen supply store (left) to be conceptually equivalent to the kitchen. Alternatively, if scenes are categorized from the similarity of spatial layout and surfaces (13–15), then observers might place the laundry room (center) into the same category as the kitchen. However, most of us share the intuition that the medieval kitchen (right) is in the same category, despite sharing few objects and features with the top image. Why is the image on the right a better category match to the modern kitchen than the other two?

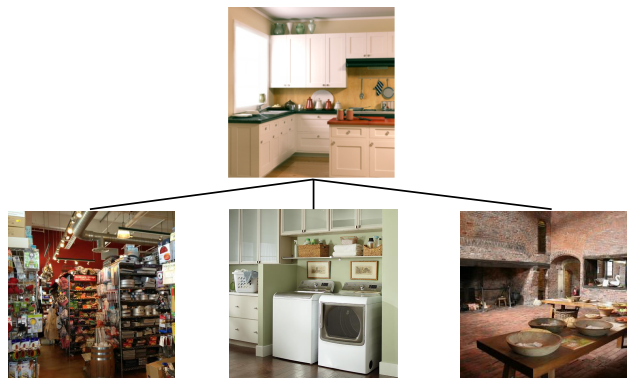

Figure 1: Which of the bottom images is in the same category as the kitchen image shown on top? Many influential models of visual perception would assume that scenes containing similar objects, such as the kitchen supply store (left), or similar layout, such as the laundry room (middle) would be placed into the same category by human observers. However, human

**observers tend to pick the medieval kitchen as the other category member despite having very different objects and features from the top kitchen.**

Here we put forth the hypothesis that the conceptual structure of environments is driven primarily by the *actions* that a scene affords. We assert that representing a scene in terms of its high-level affordances provides a better match to patterns human scene categorization than state-of-the-art models representing a scene's visual features or objects.

Figure 2 illustrates our approach. We constructed a large-scale scene category similarity matrix by querying over 2,000 observers on over 63,000 images from 1055 scene categories (Figure 2A). We compared this human response pattern with an affordance-based similarity pattern created by asking hundreds of observers to indicate which of several hundred actions could take place in each scene (Figure 2B). We found a striking resemblance between affordance-based scene similarity and the human similarity pattern. The affordance model not only explained more variance in the human category pattern than leading models of visual features and objects, but also contributed the most uniquely explained variance of any model, These results suggest that a scene's affordances provide a fundamental coding scheme for human scene categorization.

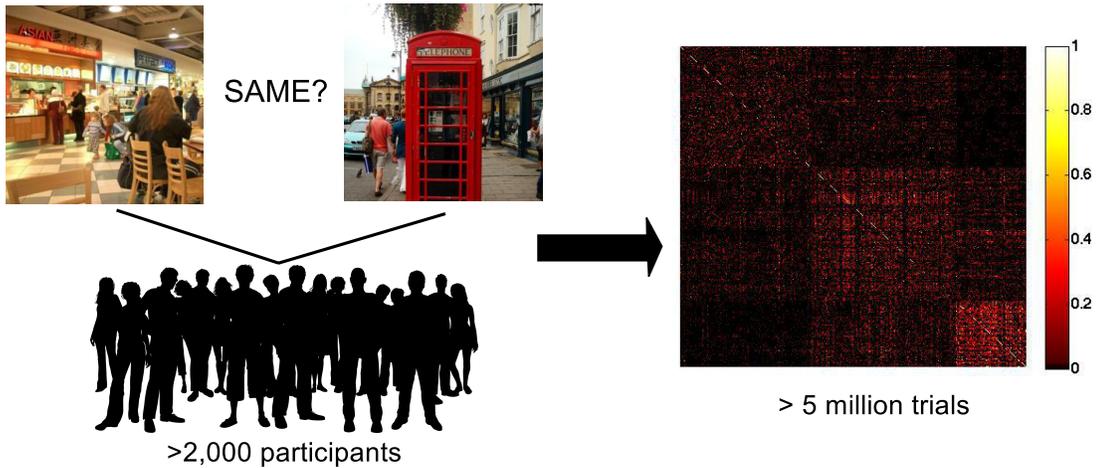

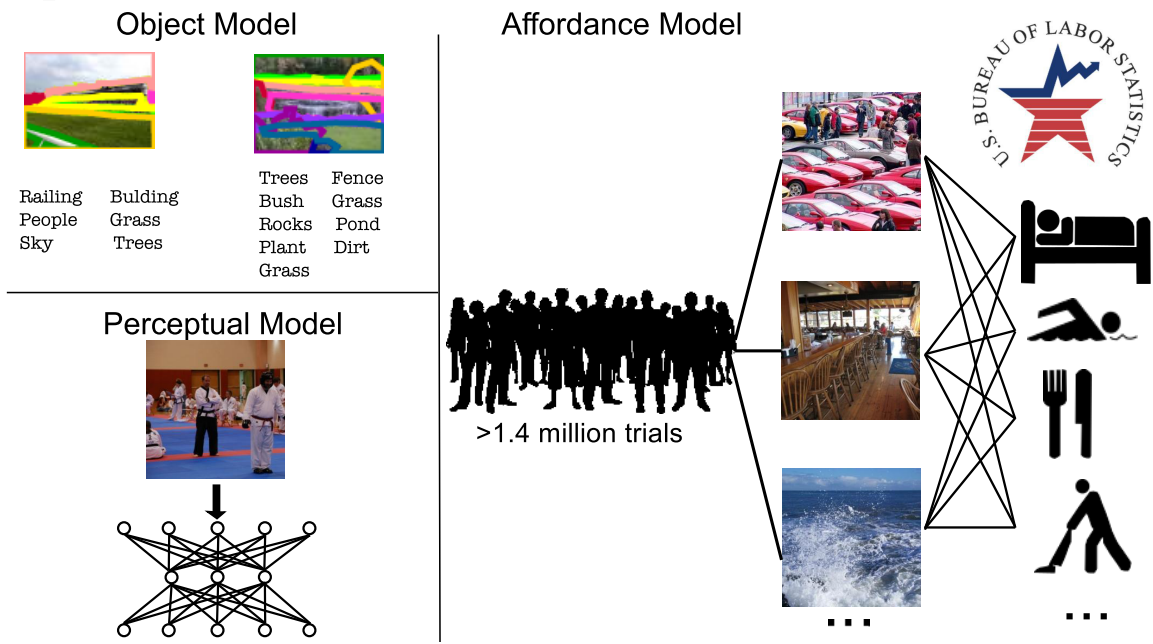

Figure 2: (A) We used a large-scale online experiment to generate a similarity matrix of scene categories. Over 2,000 individuals viewed more than 5 million trials in which participants viewed two images and indicated whether they would place the images into the same category. (B) Using the LabelMe tool (16) we examined the extent to which scene category similarity was related to having similar objects. Our perceptual model used the output features of a state-of-the-art convolutional neural network (17), to examine the extent to which low-level visual features contribute to scene category. To generate the affordance model, we took 227 actions from the American Time Use Survey. Using crowdsourcing, participants indicated which actions could be performed in which scene categories.

# Results

## *Human Scene Similarity*

To assess the conceptual structure of scene environments, we asked over 2,000 human observers to categorize images belonging to 311 scene categories in a large-scale online experiment (see Methods). These categories were chosen from a larger set of over 1,000 putative categories identified from existing scene databases and from literature review. Although category assessments were collected for all categories, we are restricting analysis to those with the highest agreement among observers (see Methods). In the experiment, observers viewed two images, either drawn from the same putative category or two randomly selected categories, and then indicated whether the two images were from the same category. The resulting 311 by 311 human distance matrix is shown in Figure 3. In order to visualize the category structure, we have ordered the scenes using the optimal leaf ordering for hierarchical clustering (18); allowing us to see what data-driven clusters emerge.

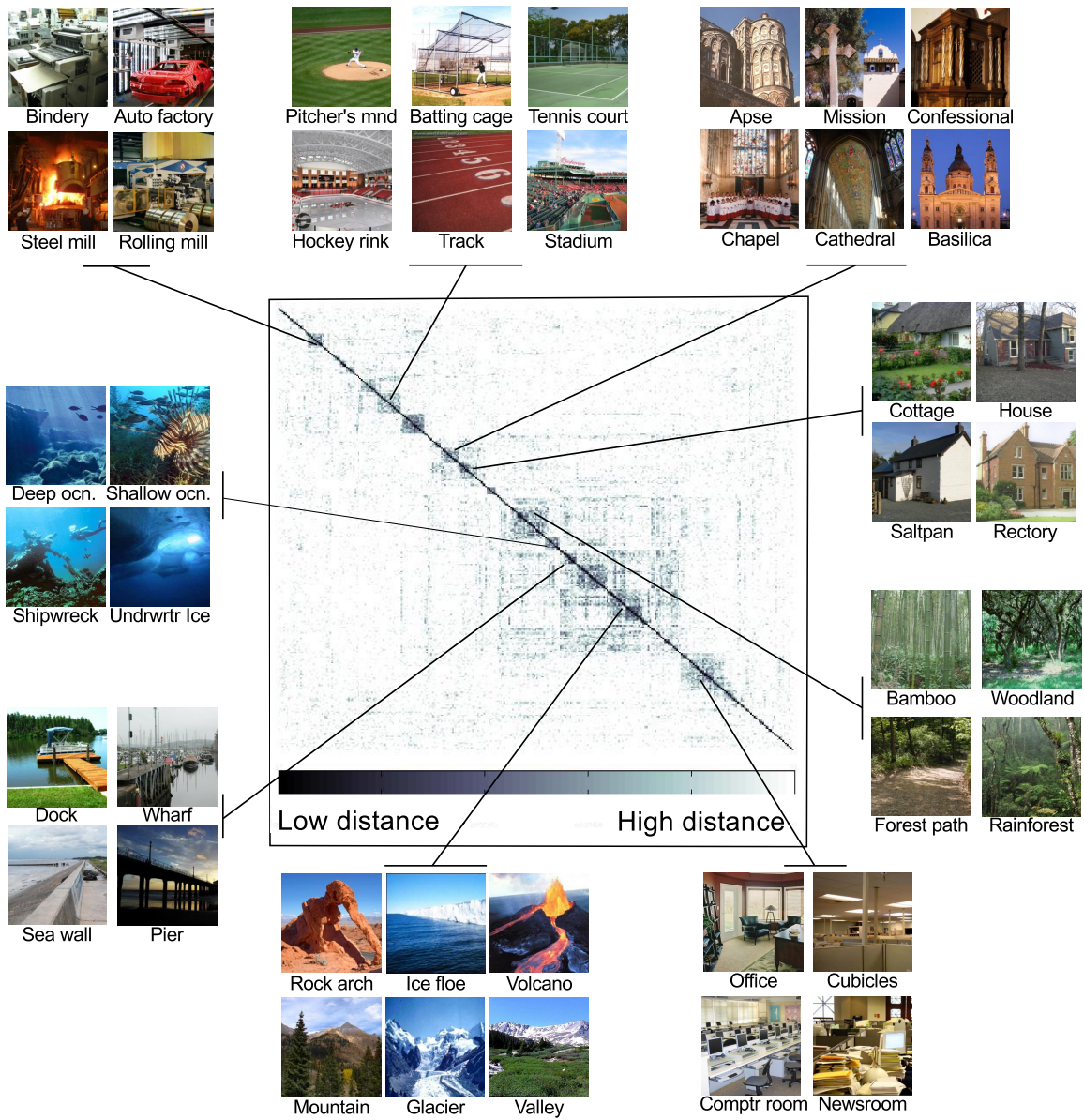

Figure 3: The human similarity matrix from our large-scale online experiment was found to be sparse. Over 2,000 individual observers categorized images in 311 scene categories. We visualized the structure of this data using optimal leaf ordering for hierarchical clustering, and show representative images from categories in each cluster.

Several category clusters are visible. Some clusters appear to group several subordinate-level categories into a single entry-level concept, such as "bamboo forest", "woodland" and "rainforest" being examples of *forests.* Other clusters seem to reflect broad classes of activities (such as "sports") which are visually heterogeneous and cross other previously defined scene boundaries, such as indoor-outdoor (10, 19–21), or the size of the space (8, 13, 22). Such activity-oriented clusters hint that the actions that one can perform in a scene (the scene's affordances) could provide a fundamental grouping principle for scene similarity.

### *Affordance-based Similarity Best Correlates with Human Similarity Structure*

We first created an affordance-based distance space from 227 actions taken from the American Time Use Survey lexicon (http://bls.gov/tus, see Methods for details). As this lexicon serves to catalog the actions taken by people in their daily lives, we reasoned that these labels provide a comprehensive space of actions that could be performed in each of our 311 scene categories. We conducted a second large-scale online experiment in which observers indicated which of the 227 could take place in each of the 311 scene categories (see Methods).

Of course, being able to perform similar actions often means manipulating similar objects, and scenes with similar objects are likely to share visual features. Therefore, we compare affordance-based categorization patterns to alternative models based on perceptual features, object-based similarity, and the semantic similarity of category names.

In order to quantify the performance of each of our models, we defined a noise ceiling based on the inter-observer reliability in the human scene distance matrix. This provides an estimate of the explainable variance in the scene categorization data, and thus provides an upper bound on the performance of any of our models. Using bootstrap sampling (see Methods), we found an inter-observer correlation of r=0.76. In other words, we cannot expect a correlation with any model to exceed this value.

Affordance-based similarity had the highest resemblance to the human similarity pattern (r=0.50). This represents about 2/3 of the maximum observable correlation obtained from the noise ceiling. As shown in Figure 4, this correlation is substantially higher than any of the alternative models we tested.

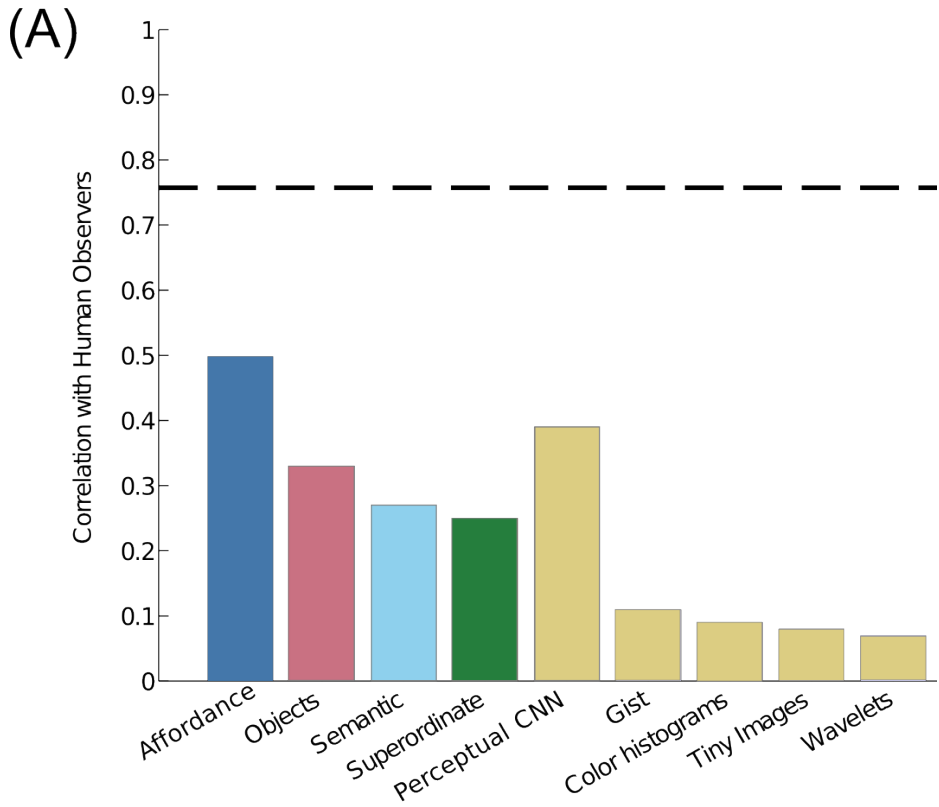
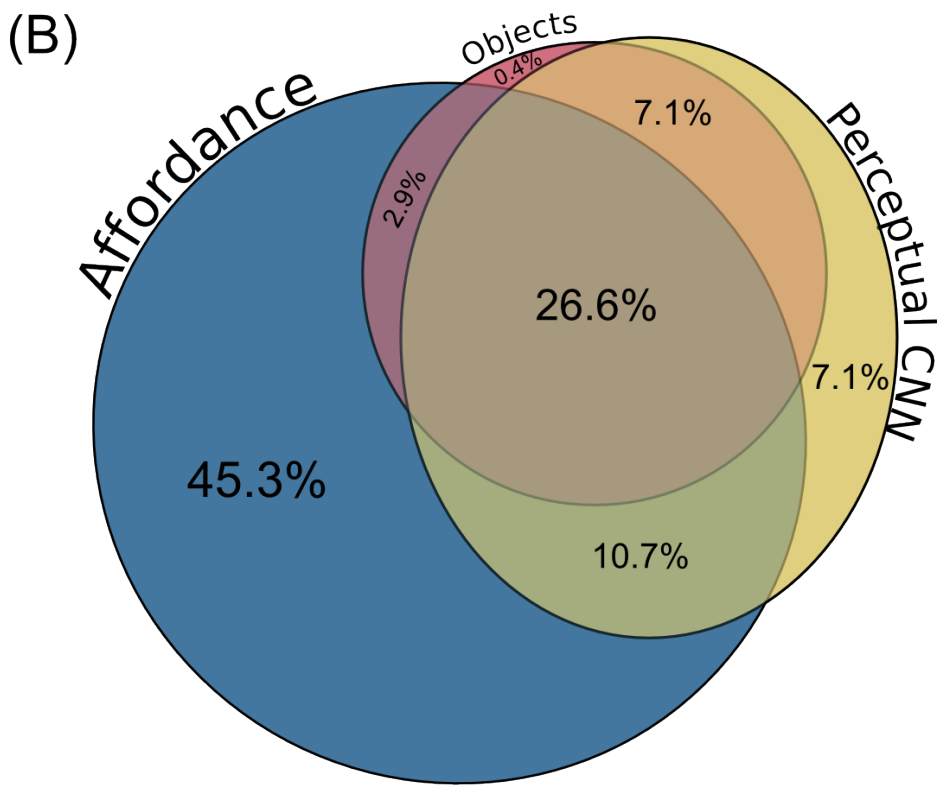

**Figure 3: (A)** Correlation of all models with human scene categorization pattern. Affordance-based similarity (dark blue, left) showed the highest resemblance to human behavior, achieving 2/3 of the maximum explainable similarity (black dotted line). Of the models based on visual features (yellow, right), only the model using the top-level features of the convolutional neural network (CNN) showed substantial resemblance to human data. Object-based similarity, semantic similarity and superordinate-level similarity all showed moderate correlations. **(B)** Euler diagram showing the distribution of explained variance for the three top-performing models. Affordance-based similarity independently explained 13.2% of the variance in the human similarity pattern (45% of total variance explained by all models). By contrast, perceptual similarity independently accounted for only 2% of the variance (7% of explained variance) and object-based similarity only accounted for 0.11% of the variance (0.4% of the explained variance).

We tested five different models based on purely visual features. The most sophisticated used the top-level features of a state-of-the-art convolutional neural network model (CNN, (17) trained on the ImageNet database (23). These features, computed by iteratively applying learned nonlinear filters to the image, have been shown to be a powerful image representation for a wide variety of visual tasks (24). Category distances in CNN space produced a correlation with human similarity of r=0.39. Simpler visual features, however, such as gist (13), color histograms (25), Tiny Images (14), and wavelets (26) had low correlations with human similarity.

Scene similarity could also be predicted to some extent based on the similarity between the objects present in scene images (r=0.33, using human-labeled objects from the LabelMe database, (16), or the semantic distance between category names in the WordNet tree (27–29)(r=0.27). Surprisingly, a model that merely groups scenes by superordinate-level categories (indoor, urban or natural environments) also had a substantial correlation (r=0.25) with human similarity patterns.

Although each of these feature spaces had differing dimensionalities, this pattern of results also holds if the number of dimensions is equalized through dimensionality reduction (see Methods and Figure S2).

### *Independent Contributions from Alternative Models*

To what extent does affordance-based similarity *uniquely* explain the patterns of human similarity? Although affordance-based similarity was the best explanation of the human similarity pattern of the models we tested, perceptual and object-based similarities also had sizeable correlations with human behavior. To what extent do these models make the same predictions?

In order to assess the independent contributions made by each of the models, we used a hierarchical linear regression analysis in which each of the three top-performing models was used either separately or in combination to predict the human similarity pattern. By comparing the $r^2$ values from the individual models to the $r^2$ values for the combined model, we can assess the unique variance explained by each descriptor. A combined model with all nine features explained 29.8% of the variance in the human similarity pattern (r=0.55). This model is driven almost entirely by the top three feature spaces (affordance, perceptual CNN, and object labels), which explained a combined 29.1% of the variance (r=0.54). Note that affordances explained 85.6% of this explained variance, indicating that the object and perceptual features only added a small amount of independent information (14.4% of the combined variance).

Although there was a sizable overlap between the portions of the variance explained by each of the models (see Figure 4B), nearly half of the total variance explained can be attributed only to affordances (13.2% of total variance, or 45.3% of the explained variance), and was not shared by the other two models. In contrast, the independent variance explained by perceptual similarity and object-based similarity accounted for only 2% (7% of explained variance) and 0.11% (0.4% of explained variance) of the total variance respectively. Therefore, the contributions of perceptual and object-based similarities are largely shared with affordance-based similarity, further highlighting the utility of affordances for explaining human scene similarity patterns.

### *Examining affordance space*

In order to better understand the affordance space, we performed classical multi-dimensional scaling on the affordance distance matrix, allowing us to identify how patterns of affordances contribute to the overall similarity pattern. We found that at least 10 MDS dimensions were necessary to explain 95% of the variance in the affordance distance matrix, suggesting that the efficacy of the affordance-based model was driven by a number of distinct affordance dimensions. We examined the projection of categories onto the first three MDS dimensions. As shown in Figure 5, the first dimension appears to separate indoor locations that have a high potential for social interactions (such as "socializing" and "attending meetings for personal interest") from outdoor spaces that afford more solitary activities, such as "hiking" and "science work". The second dimension separates work from leisure. Later dimensions appear to separate environments related to transportation and industrial workspaces from restaurants, farming, and other food-related environments (see Figure S1).

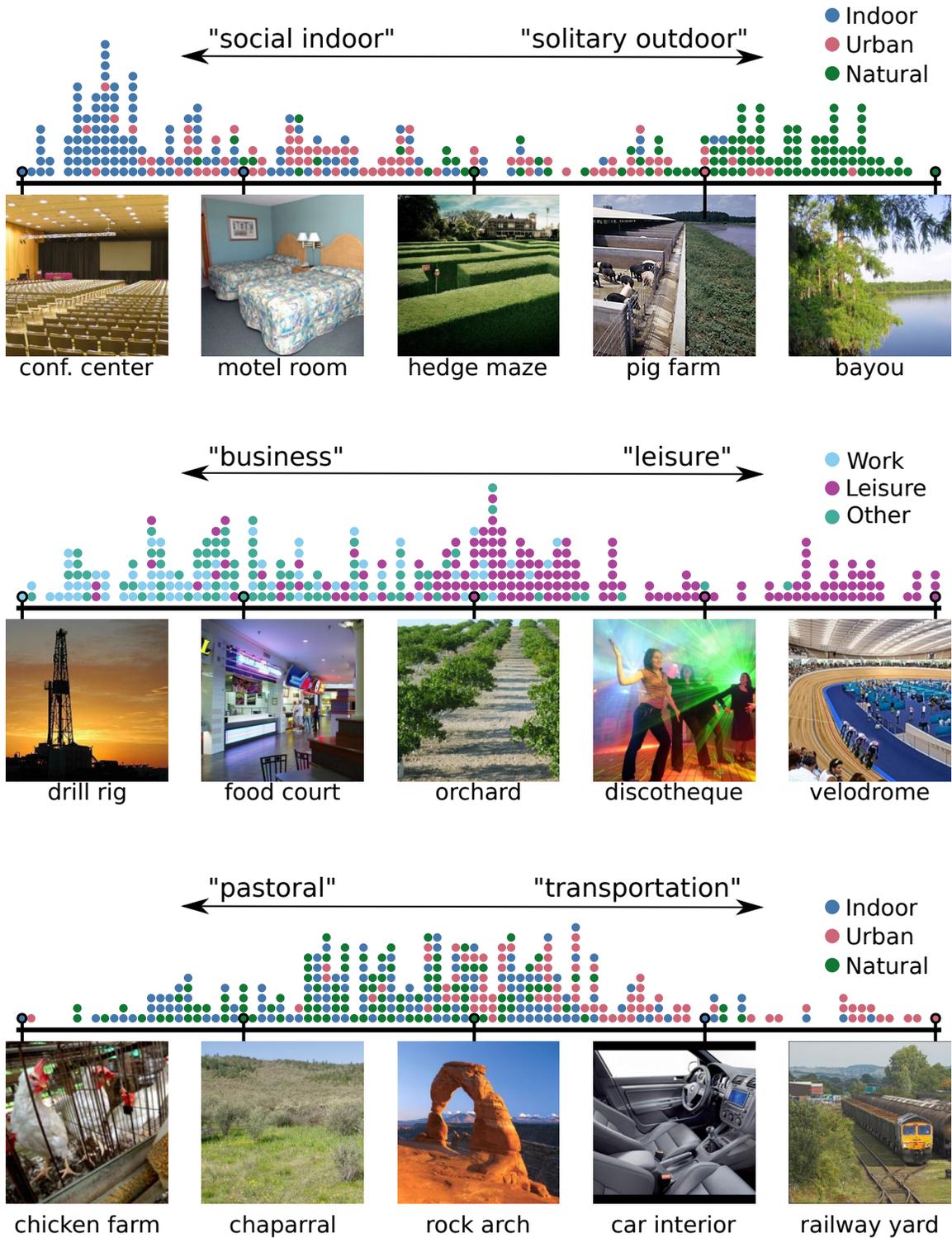

**Figure 4: (Top):** Distribution of superordinate-level scene categories along the first MDS dimension of the affordance distance matrix, which separates indoor scenes from natural scenes. Actions that were positively correlated with this

**component tend to be outdoor-related activities such as *hiking* while negatively correlated actions tend to reflect social activities such as *eating and drinking*. (Middle) The second dimension seems to distinguish environments for work from environments for leisure. Actions such as *playing games* are positively correlated while actions such as *construction and extraction work* are negatively correlated (Bottom). The third dimension distinguishes environments related to farming and food production (pastoral) from industrial scenes specifically related to transportation. Actions such as *travel* and *vehicle repair* are highly correlated with this dimension, while actions such as *farming* and *food preparation* are most negatively correlated.**

**Discussion**

We have shown that human scene categorization is better explained by the action possibilities, or affordances, of a scene than by the scene's visual features or objects. Furthermore, affordance-based similarity explained far more independent variance than did alternative models, as these models were correlated with human similarity only insofar as they were also correlated with the scene's affordances. This suggests that a scene's affordances contain essential information for categorization that is not captured by the scene's objects or low-level visual features.

The current results cannot be explained by the smaller dimensionality of the affordance-based features, as further analysis revealed that affordance-based features outperformed other similarity spaces using equivalent numbers of dimensions. Furthermore, this pattern was observed over a wide range of dimensions, suggesting that each affordance feature contained more information about scene similarity than each perceptual or object-based feature.

Our current results reflect the first large-scale operationalization of J.J. Gibson's influential theory of ecological visual perception in which he asserted that vision operates through the direct perception of an environment's affordances. Although the particulars of the theory have been controversial (30), previous small-scale studies have found that environmental affordances such as navigability are reflected in patterns of human categorization (8, 31), and are perceived very rapidly from images (9). The current results show that this principle is true for scene perception generally, holding true over a comprehensive set of hundreds of scene image categories.

The idea that vision functions for action has permeated the literature of visual perception, but it has been difficult to fully operationalize this idea for testing. Psychologists have long theorized that rapid and accurate environmental perception could be achieved by the explicit coding of an environment's affordances, most notably in J.J. Gibson's influential theory of ecological perception (32). Although this work is most often remembered as a theory of object affordances (a chair is an object that affords sitting), Gibson's theory aimed to explain affordances at the scale of the environment, with objects being one environmental element, along with land, water, sky and other animals. Our results provide the first comprehensive, data-driven test of this hypothesis, using data from hundreds of scene categories and affordances. By leveraging the power of crowdsourcing, we were able to obtain both a large-scale similarity structure for visual scenes, but also normative ratings of affordances for these scenes. Using hundreds of categories, thousands of observers and millions of observations, crowdsourcing allowed a scale of research previously unattainable. Previous research on scene affordance has also suffered from the lack of a comprehensive list of affordances, relying instead on the free responses of human observers describing the actions that could be taken in scenes (8, 33). By using an already comprehensive set of actions from the American Time Use Survey, we were able to see the full power of affordances for predicting human categorization patterns.

Given the relatively large proportion of variance independently explained by affordance-based similarity, we are left with the question of why this model outperforms the more classic models. By examining patterns of variance in the affordance by category matrix, we found that affordances can be used to separate scenes along previously defined dimensions of scene variance, such as superordinate-level category (21, 34, 35), and between work and leisure activities (36). Although the variance explained by affordance-based similarity does not come directly from visual features or the scene's objects, human observers must be able to apprehend these affordances from the image somehow. It is therefore a question open for future work to understand the extent to which human observers bring non-visual knowledge to bear on this problem.

Some recent work has examined large-scale neural selectivity based on semantic similarity (29), or object-based similarity (7), finding that both types of conceptual structures can be found in the large-scale organization of human cortex. Our current work indeed shows sizeable correlations between these types of similarity structures and human behavioral similarity. However, we find that affordance-based similarity is a better predictor of behavior and may provide an even stronger grouping principle in the brain.

These results challenge many existing models of visual categorization that consider categories to be purely a function of shared visual features or objects. Just as the Aristotelian theory of concepts assumed that categories could be defined in terms of necessary and sufficient features, classical models of visual categorization have assumed that a scene category can be explained by necessary and sufficient objects (2, 7) or diagnostic visual features (37, 38). However, just as the classical theory of concepts cannot account for important cognitive phenomena, the classical theory of scene categories cannot account for the fact that two scenes can share a category even when they do not share many features or objects. By contrast, the current results demonstrate that the possibility for action creates categories of environmental scenes. In other words, a kitchen is a kitchen because it is a space that affords cooking, not because it shares objects or other visual features with other kitchens.

**Methods**

*Creating Human Scene Similarity Matrix*

We wanted to amass a comprehensive collection of scene categories with high human participant agreement about membership. We started with 1,055 scene categories identified from the SUN and ImageNet databases (23, 39) and from literature review. These databases used the WordNet (27) hierarchy to identify potential scene concepts. We only included categories with at least 20 image exemplars, for a grand total of 63,988 images.

Human scene similarity was assessed using a large-scale online study using Amazon's Mechanical Turk. Potential participants were recruited from a pool of trusted observers with at least 2,000 previously approved trials with at least 98% approval. Additionally, participants were required to pass a brief scene vocabulary test before participating.

Each trial consisted of 10 sub-trials in which two images were presented side by side. Half of the image pairs came from the same putative scene category, while the other half were from two different categories randomly selected. Image exemplars were randomly selected within a category on each trial. Participants were instructed to indicate whether they would put the two images into the same category, and to type in the category name they would use for the left image (not analyzed, but used to assess understanding of the task). Workers were compensated $0.02 for each trial. We obtained 10 independent observations for each cell in the 1055 by 1055 scene matrix, for a total of over 5 million trials. Individual participants completed a median of 5 hits of this task (range: 1-36,497). There was a median of 1,116 trials in each of the diagonal entries of the matrix, and a median of 11 trials in each cell of the off-diagonal entries.

From the 1,055 by 1,055 category similarity matrix, we identified 311 categories with the strongest within-category cohesion (at least 7 of the 10 observers agreed that images were from the same category). Thus, 885,968 total trials were included from 2,296 individual workers in the final dataset.

### *Creating the Scene Affordance Space*

In order to determine whether scene categories are governed by affordance similarity, we needed a broad space of possible actions that could take place in scenes. We gathered these actions from the American Time Use Survey (ATUS), sponsored by the US Bureau of Labor Statistics, and representing United States census data. The lexicon used in this study was pilot tested over the course of three years (40). The ATUS lexicon includes 428 specific activities organized into 17 major activity categories and 105 mid-level categories. The 227 actions included in our study included the most specific category levels with the following exceptions:

(1) The superordinate category "Caring for and Helping Non-household members" was dropped as these actions would be visually identical to those in the "Caring for and Helping Household members" category.

(2) In the ATUS lexicon, the superordinate-level category "Work" contained only two specific categories (primary and secondary jobs). Because different types of work can look very visually different, we expanded this category by adding 22 categories representing the major labor sectors from the Bureau of Labor Statistics.

(3) The superordinate-level category "Telephone calls" was collapsed into one action because we reasoned that all telephone calls would look visually similar.

(4) The superordinate-level category "Traveling" was similarly collapsed into one category because being in transit to go to school should be visually indistinguishable from being in transit to go to the doctor.

(5) All instances of "Security procedures" have been unified under one category for similar reasons.

(6) All instances of "Waiting" have been unified under one category.

(7) All "Not otherwise specified" categories have been removed.

The final list of actions can be found in the Supplemental Materials.

### Action Space Norming Method

Using a separate large-scale online experiment, 484 workers indicated which of the 227 actions could take place in each of the 311 scene categories. Participants were screened using the same criterion described above. In each trial, a participant would see a randomly selected exemplar image of a scene category along with a random selection of 17 or 18 of the 227 actions. Each action was hyperlinked to its description in the ATUS lexicon. Participants were instructed to check which of the actions would typically be done in the type of scene shown.

Each individual participant did a median of 9 trials (range: 1-4,868). Each scene category – action pair was rated by a median of 16 participants (range: 4-86), for a total of 1.2 million trials.

We created a 311-category by 227-matrix in which each cell represents the proportion of participants indicating that the action could take place in the category. Since scene categories vary widely in the number of actions they afford, we obtained a distance matrix by computing the cosine distance between categories; this measures the overlap between affordances while being invariant to the absolute magnitude of the action vector.

### Affordance MDS Analysis

To better understand the affordance dimensions that give rise to scene similarities, we performed a classical multidimensional scaling (MDS) decomposition of the action distance matrix. This yielded an embedding of the scene categories such that inner products in this embedding space approximate the (double-centered) distances between scene categories, with the embedding dimensions ranked in order of importance (41). In order to associate affordances with each of these dimensions, we computed the correlation coefficient between each action (across scene categories) with the category coordinates for a given dimension.

### *Alternative Models*

To put the performance of the action-based model in perspective, we compared it to eight other models. Five of the models represented visual features, and one model examined the objects that were present in the scenes. These models yielded scene category by feature matrices, and were converted to distance matrices using cosine distance. Additionally, two models measured distances directly, based either on the semantic distance between scene category names, or simply by whether scenes belonged to the same superordinate level category (indoor, urban or natural). We will detail each of the models below.

### *Perceptual Models*

### Convolutional Neural Network

We generated a perceptual feature vector using the publicly distributed OverFeat convolutional neural network (CNN) (17), which was trained on the ImageNet 2012 training set (23). This 7-layer CNN takes an image of size 231x231 as input, and produces a vector of 4096 image features that are optimized for 1000-way object classification. This network achieves top-5 object recognition on ImageNet 2012 with approximately 16% error, meaning that the correct object is one of the model's first five guesses in 84% of trials. Using the top layer of features, we averaged the features for all images in each scene category to create a 311-category by 4096-feature matrix.

### Gist

We used the Gist descriptor features of (13). This popular model for scene recognition provides a summary statistic representation of the dominant orientations and spatial frequencies at multiple scales coarsely localized on the image plane. We used spatial bins at 4 cycles per image and 8 orientations at each of 4 spatial scales for a total of 3,072 filter outputs per image. We averaged the gist descriptors for each image in each of the 311 categories to come up with a single 3072-dimensional descriptor per category.

### Color histograms

We represented color using LAB color space. For each image, we created a two-dimensional histogram of the a* and b* channels using 50 bins per channel. We then averaged these histograms over each exemplar in each category, such that each category was represented as a 2500 length vector representing the averaged colors for images in that category. The number of bins was chosen to be similar to those used in previous scene perception literature (25).

### Tiny Images

Torralba and colleagues (14) demonstrated that human scene perception is robust to aggressive image downsampling, and that an image descriptor representing pixel values from such downsampled images could yield good results in scene classification. Here, we downsampled each image to 32 by 32 pixels (grayscale). We created our 311-category by 1024 feature matrix by averaging the downsampled exemplars of each category together.

### *Wavelets*

We represented each image in this database as the output of a bank of multi-scale Gabor filters. This type of representation has been used to successfully model the representation in early visual areas (26). Each image was converted to grayscale, down sampled to 128 by 128 pixels, and represented with a bank of Gabor filters at three spatial scales (3, 6 and 11 cycles per image with a luminance-only wavelet that covers the entire image), four orientations (0, 45, 90 and 135 degrees) and two quadrature phases (0 and 90 degrees). An isotropic Gaussian mask was used for each wavelet, with its size relative to spatial frequency such that each wavelet has a spatial frequency bandwidth of 1 octave and an orientation bandwidth of 41 degrees. Wavelets were truncated to lie within the borders of the image. Thus, each image is represented by $3*3*2*4+6*6*2*4+11*11*2*4 = 1328$ total Gabor wavelets. We created the feature matrix by averaging the Gabor weights over each exemplar in each category.

### *Object-based Model*

In order to model the similarity of objects within scenes, we employed the LabelMe tool (Russell et al, 2008) that allows users to outline and annotate each object in each image by hand. 7,710 scenes from our categories were already labeled in the SUN 2012 release (39), and we augmented this set by labeling an additional 223 images. There were a total of 3,563 unique objects in this set. Our feature matrix consisted of the proportion of scene images in each category containing a particular object. For example, if 10 out of 100 *kitchen* scenes contained a "blender", the entry for kitchen-blender would be 0.10. In order to estimate how many labeled images we would need to represent a scene category, we performed a bootstrap analysis in which we resampled the images in each category with replacement

(giving the same number of images per category as in the original analysis), and then measured the variance in distance between categories. With the addition of our extra images, we ensured that all image categories either had at least 10 labeled images or had mean standard deviation in distance to all other categories of less than 0.05 (e.g. less than 5% of the maximal distance value of 1).

### *Semantic Models*

We examined semantic similarity by examining the shortest path between category names in the WordNet tree using the implementation of (28). The similarity matrix was normalized and converted into distance. We examined each of the metrics of semantic relatedness implemented in Wordnet::Similarity and found that this path measure was the best correlated with human performance.

### *Superordinate-Category Model*

As a baseline model, we examined how well a model that groups scenes only according to superordinate-level category would predict human scene similarity assessment. We assigned each of the 311 scene categories to one of three groups (natural outdoors, urban outdoors or indoor scenes). Then, each pair of scene categories in the same group was given a distance of 0 while pairs of categories in different groups were given a distance of 1.

### *Noise Ceiling*

The variability of human similarity responses puts a limit on the maximum correlation expected by any of the tested models. In order to get an estimate of this maximum correlation, we used a bootstrap analysis in which we sampled with replacement observations from our dataset to create two new datasets of the same size as our original dataset. We the correlated these two datasets to one another, and repeated this process 1000 times.

### *Regression Analysis*

In order to understand the unique variance contributed by each of our feature spaces, we used hierarchical linear regression analysis, using each of the feature spaces both alone and in combination to predict the human similarity response pattern. In total, eight regression models were used: (1) all nine feature spaces used together; (2) the top 3 performing features together (affordances, objects and the perceptual CNN); (3-5) each of the top three features alone; (6-8) each pair of the top three features. By comparing the $r^2$ values of a feature space used alone to the $r^2$ values of that space in conjunction with another feature space, we can infer the amount of variance that is independently explained by that feature space. In order to visualize this information in an Euler diagram, we used EulerAPE software (42).

**Supplemental Information**

Supplementary Figures

|  Scene Scores | | Correlated Actions | |
| --- | --- | --- | --- |
| Highest | Lowest | Highest | Lowest |

### Dim 1: 45%

| Highest | Lowest | Highest | Lowest |
| --- | --- | --- | --- |
| bayou | conference center | Farming / Fishing and Forestry work | Arts / Design / Entertainment / Sports / Media work |
| swamp | music store | Travel | Socializing |
| chaparral | piano store | Hiking | Volunteer at event |
| forest | conference hall | Science work | Eating & drinking |
| waterfall | discotheque | Fishing | Sales work |
| waterfall (cascade) | video store | Camping | Attending school-related meetings & conferences |
| ice shelf | music studio | Rock climbing / caving | Attending meetings for personal interest |
| rainforest | theater (indoor round) | Watching fishing | Volunteer work: fundraising |
| sea cliff | movie theatre (indoor) | Hunting | Listening to music (not radio) |
| wheat field | bakery | Walking | Community and Social work |

### Dim 2: 20%

| Highest | Lowest | Highest | Lowest |
| --- | --- | --- | --- |
| wrestling ring | drill rig | Arts / Design / Entertainment / Sports / Media work | Income-generating services |
| bullpen | drugstore | Work-related sports | Architecture and Engineering work |
| velodrome (indoor) | auto factory | Playing games | Construction and Extraction work |
| batting cage (indoor) | call center | Playing sports with children | Sales work |
| batting cage (outdoor) | cubicle | Extracurricular club activities | Work-related eating/drinking |
| aquatic theater | control tower (indoor) | Hobbies | Job search activities |
| arena (basketball) | office cubicles | Watching weightlifting | Business and Financial Operations work |
| bullring | kitchenette | Playing basketball | Homework |
| stadium | pharmacy | Attending child's events | Food & drink preparation |
| track | chemical plant | Working out | Interior decoration & repair |

### Dim 3: 9%

| Highest | Lowest | Highest | Lowest |
| --- | --- | --- | --- |
| railway yard | indoor chicken farm | In transit / traveling | Farming / Fishing and Forestry work |
| tunnel (road outdoor) | pig farm | Transportation and Material Moving work | Food & drink preparation |
| arrival gate | vegetable garden | Architecture and Engineering work | Food Preparation and Serving work |
| access road | hayfield | Installation / Maintenance and Repair work | Food presentation |
| highway | dairy (indoor) | Travel | Eating & drinking |
| tunnel (rail outdoor) | corn field | Construction and Extraction work | Purchasing food (not groceries) |
| subway interior | pantry | Using vehicle maintenance & repair services | Volunteer work: food preparation |
| truss bridge | bakery | Security screening | Exercising & playing with animals |
| tollbooth | chicken yard | Vehicle repair & maintenance (self) | Kitchen & food clean-up |
| heliport | delicatessen | Attending museums | Using meal preparation services |

### Dim 4: 7%

| Highest | Lowest | Highest | Lowest |
| --- | --- | --- | --- |
| pump room | dining car | Architecture and Engineering work | In transit / traveling |
| nuclear power plant (indoor) | car interior | Construction and Extraction work | Eating & drinking |
| particle accelerator | bus interior | Production work | Travel |
| power plant (indoor) | airplane cabin | Installation / Maintenance and Repair work | Food Preparation and Serving work |
| water treatment plant (indoor) | limousine interior | Science work | Food presentation |
| bindery | restaurant | Computer and Mathematical work | Purchasing food (not groceries) |
| oil refinery | ice cream shop | Business and Financial Operations work | Food & drink preparation |
| machine shop | liquor store (indoor) | Using home repair & construction services | Transportation and Material Moving work |
| electrical substation | bus depot | Building and Grounds Cleaning and Maintenance work | Grocery shopping |
| rolling mill | subway interior | Income-generating services | Using meal preparation services |

**Supplementary Figure 1: Principal components of action matrix.** MDS was performed on the scene by action matrix, yielding a coordinate for each scene along each MDS dimension, as well as a correlation between each action and each dimension.

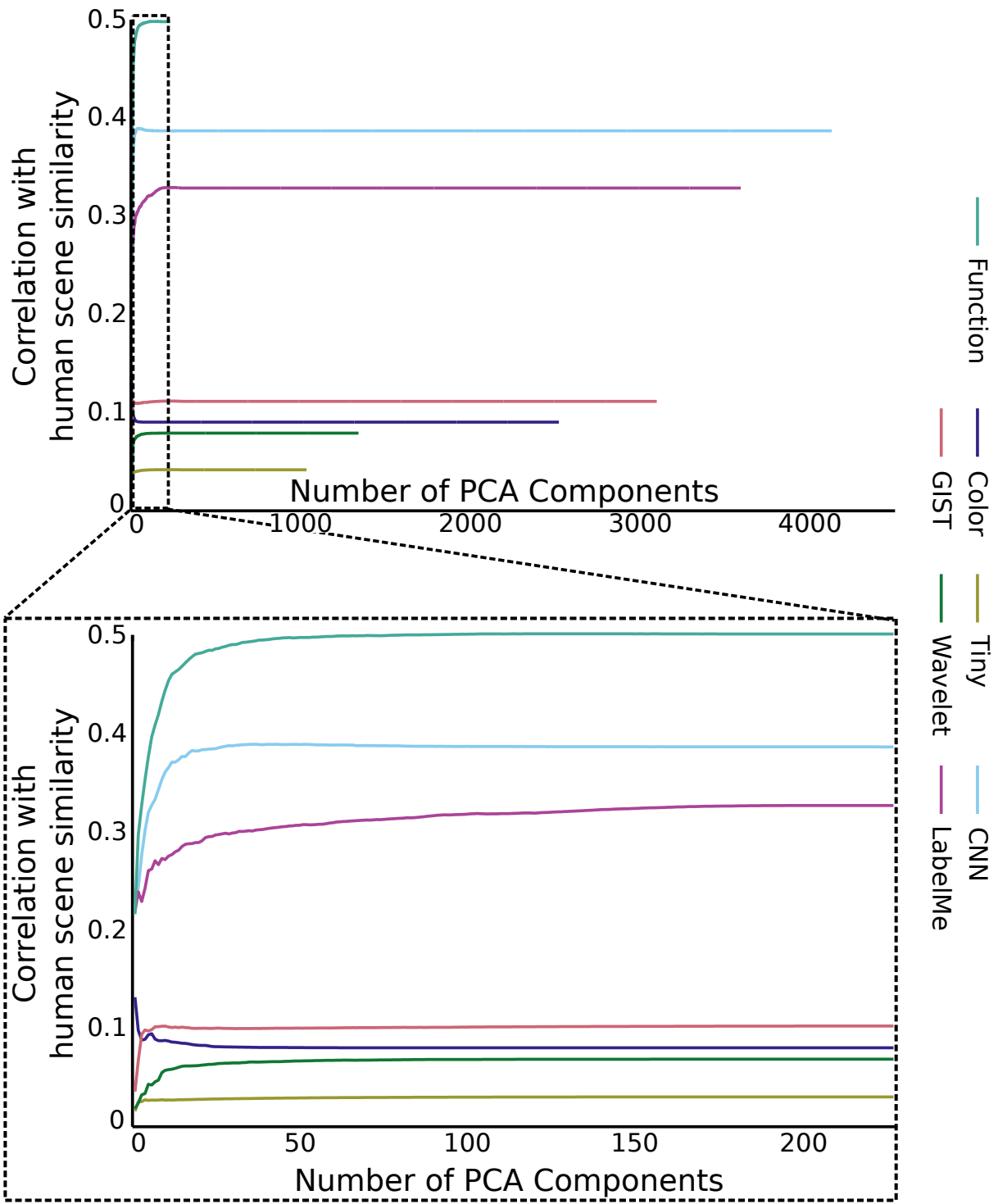

**Supplementary Figure 2: Robustness to dimensionality reduction.** For each feature space, we reconstructed the feature matrix using a variable number of PCA components and then correlated the cosine distance in this feature space with the human scene distances. Although the number of features varies widely between spaces, all can be described in ~100 dimensions, and the ordering of how well the features predict human responses is essentially the same regardless of the number of dimensions.

**List of Affordances**

I. Personal care
   Health related self-care
   Sexual activity
   Sleeping
   Washing/dressing/grooming oneself

II. Household activities
   Appliance repair & maintenance (self)
   Building & repairing furniture
   Cleaning home exterior
   Email
   Exercising & playing with animals
   Exterior home repair & decoration
   Financial management
   Food & drink preparation
   Food presentation
   Grocery shopping
   Home heating / cooling
   Home security
   Home-schooling children
   Household organization & planning
   Interior decoration & repair
   Interior home cleaning
   Kitchen & food clean-up
   Laundry
   Lawn/garden & plant care
   Mailing
   Maintaining home pool/pond/hot tub
   Non-veterinary pet care
   Sewing & repairing textiles
   Storing household items
   Vehicle repair & maintenance (self)

III. Caring for & helping household members
   Arts & crafts with children
   Attending child's events
   Helping adult
   Helping child with homework
   Looking after adult
   Looking after children
   Obtaining medical care for adult
   Obtaining medical care for child

        Organizing & planning for adults
        Organizing & planning for children
        Physical care of adult
        Physical care of children
        Picking up / dropping off adult
        Picking up / dropping off child
        Playing sports with children
        Playing with children (not sports)
        Providing medical care to adult
        Providing medical care to child
        Reading with children
        Talking with children

IV.     Work & work-related activities
        Architecture & engineering work
        Arts / Design / Entertainment / Sports / Media work
        Building and Grounds Cleaning and Maintenance work
        Business and Financial Operations work
        Community and social work
        Computer and mathematical work
        Construction and Extraction work
        Education and library work
        Farming / Fishing and Forestry work
        Food Preparation and Serving work
        Healthcare work
        Income-generating hobbies & crafts
        Income-generating performance
        Income-generating rental property activity
        Income-generating selling activities
        Income-generating services
        Installation / Maintenance and Repair work
        Job interviewing
        Job search activities
        Legal work
        Management/Executive work
        Military work
        Office and Administrative work
        Personal Care and Service work
        Production work
        Protective services work
        Sales work
        Science work
        Transportation and Material Moving work
        Work-related eating/drinking
        Work-related social activities
        Work-related sports

V. Education
   Attending school-related meetings & conferences
   Education-related administrative activities
   Extracurricular club activities
   Homework
   School music activities
   Student government
   Taking class for degree or certification
   Taking class for personal interest

VI. Consumer purchases
   Comparison shopping
   Purchasing food (not groceries)
   Purchasing gasoline
   Shopping (except food and gas)

VII. Professional & personal care services
   Banking
   Buying & selling real estate
   Out-of-home medical services
   Using clothing repair & cleaning services
   Using legal services
   Using meal preparation services
   Using other financial services
   Using personal care services
   Using professional photography services
   Using vehicle maintenance & repair services
   Using veterinary services

VIII. Household services
   Using home repair & construction services
   Using in-home medical services
   Using interior home cleaning services
   Using lawn & garden services
   Using paid childcare services
   Using pet services

IX. Government services & civic obligations
   Civic obligations
   Obtaining licenses & paying fees
   Security screening
   Using police & fire services
   Using social services
   Waiting

X.     Eating & drinking
      Eating & drinking

XI.     Socializing, relaxing & leisure
      Arts & crafts
      Attending meetings for personal interest
      Attending movies
      Attending museums
      Attending or hosting parties
      Attending the performing arts
      Collecting as a hobby
      Computer use (not games)
      Dancing
      Gambling
      Hobbies
      Listening to music (not radio)
      Listening to radio
      Playing games
      Reading for personal interest
      Relaxing
      Socializing
      Tobacco use
      Watching television & movies
      Writing for personal interest

XII.     Sports, exercise & recreation
      Biking
      Boating
      Bowling
      Camping
      Doing aerobics
      Doing gymnastics
      Doing martial arts
      Fencing
      Fishing
      Golfing
      Hiking
      Hunting
      Participating in aquatic sports
      Participating in equestrian sports
      Participating in rodeo
      Playing baseball
      Playing basketball
      Playing billiards
      Playing football
      Playing hockey

Playing racquet sports
Playing rugby
Playing soccer
Playing softball
Playing volleyball
Rock climbing / caving
Rollerblading / skateboarding
Running
Skiing / ice skating / snowboarding
Using cardiovascular equipment
Vehicle racing/touring
Walking
Watching aerobics
Watching aquatic sports
Watching biking
Watching billiards
Watching boating
Watching bowling
Watching dance
Watching equestrian sports
Watching fencing
Watching fishing
Watching golf
Watching gymnastics
Watching hockey
Watching live baseball
Watching live basketball
Watching live football
Watching live soccer
Watching live softball
Watching live vehicle racing
Watching martial arts
Watching people walk
Watching racquet sports
Watching rock climbing / caving
Watching rodeo
Watching rollerblading / skateboarding
Watching rugby
Watching running
Watching skiing / snowboarding
Watching volleyball
Watching weightlifting
Watching wrestling
Weightlifting
Working out
Wrestling

      Yoga

XIII.     Religious & spiritual activities
      Attending religious services
      Religious education
      Religious practices

XIV.     Volunteer activities
      Volunteer at event
      Volunteer work: attending meeting
      Volunteer work: blood donation
      Volunteer work: building
      Volunteer work: clean up
      Volunteer work: collecting goods
      Volunteer work: computer use
      Volunteer work: food preparation
      Volunteer work: fundraising
      Volunteer work: organizing
      Volunteer work: performing
      Volunteer work: providing care
      Volunteer work: public safety
      Volunteer work: reading
      Volunteer work: teaching
      Volunteer work: telephone calls
      Volunteer work: writing

XV.     Telephone calls
      Telephone calls

XVI.     Traveling
      In transit / traveling
      Travel